\documentclass[aps,prl, reprint, superscriptaddress, 10pt]{revtex4-2}
\usepackage{amsmath}
\usepackage{amsfonts}
\usepackage{amssymb}
\usepackage{amsthm}
\usepackage{bm}
\usepackage{graphicx}
\usepackage{url}
\usepackage{lmodern}
\usepackage[section]{placeins}
\usepackage{xcolor}
\usepackage{multirow}
\usepackage[utf8]{inputenc}
\usepackage[T1]{fontenc}
\usepackage{hyperref}
\hypersetup{
	colorlinks = true,
	citecolor = {black}, urlcolor = {blue}, filecolor= {black}, linkcolor={black}}

\newcommand{\bi}{\begin{itemize}}
\newcommand{\ei}{\end{itemize}}

\newcommand{\bea}{\begin{eqnarray}}
\newcommand{\eea}{\end{eqnarray}}
\newcommand{\be}{\begin{equation}}
\newcommand{\ee}{\end{equation}}
\newcommand{\ben}{\begin{eqnarray*}}
\newcommand{\een}{\end{eqnarray*}}
\newcommand{\bem}{\begin{pmatrix}}
\newcommand{\eem}{\end{pmatrix}}
\newcommand{\bl}{\begin{align}}
\newcommand{\el}{\end{align}}
\newcommand{\beg}{\begin{gather}}
\newcommand{\eeg}{\end{gather}}

\newcommand{\cH}{\mathcal{H}}

\newcommand{\IH}{\mathbb{H}}

\newcommand{\g}{\gamma}

\newcommand{\m}{\mu}

\renewcommand{\t}{\tau}

\newcommand{\1}{\textbf{1}}

\newcommand{\TrH[1]}{ {\raise -.5em
	\hbox{$\buildrel {\textstyle {\rm Tr } }\over
{\scriptscriptstyle \cH _ {#1}}$}~}}
\newcommand{\res[1]}{{\raise -.5em 
\hbox{$\buildrel{\textstyle{\rm Res}}\over {\scriptscriptstyle {#1}}$}}}
\newcommand{\tends[1]}{{\raise -.5em 
\hbox{$\buildrel{\longrightarrow}\over {\scriptscriptstyle {#1}}$}}}

\newcommand{\half}{\frac{1}{2}}

\newcommand{\Tr}{\mbox{Tr}}
\newcommand{\tr}{\mbox{tr}}

\renewcommand{\Re}{\mbox{Re}}

\renewcommand{\arraystretch}{1.75}

\def\dbend{\lower3.5pt\hbox{\manual\char127}}

\def\IL{\relax{\rm I\kern-.18em L}}
\def\IH{\relax{\rm I\kern-.18em H}}

\keywords{quantum entanglement, superstrings, black holes, holography}

\begin{document}
\title{Edge Modes on Stringy Horizons}

\preprint{}

\author{Atish Dabholkar}
\affiliation{The Abdus Salam International Centre for Theoretical Physics\\
Strada Costiera 11, 34151 Trieste, Italy}

\author{Eleanor Harris}
\affiliation{The Abdus Salam International Centre for Theoretical Physics\\
Strada Costiera 11, 34151 Trieste, Italy}

\author{Upamanyu Moitra}
\affiliation{Institute for Theoretical Physics, Institute of Physics\\
Universiteit van Amsterdam, Science Park 904\\
1089 XH Amsterdam, The Netherlands}

\begin{abstract}

For a quantum field of arbitrary mass and spin in the static patch of de Sitter spacetime, the Euclidean partition function receives contributions from edge modes localized on the horizon, expressible in terms of the Harish-Chandra character of the de Sitter group. Considering the flat limit and summing over all string fields, we obtain the partition function of edge modes in string theory near the Minkowski-Rindler horizon. Application of the Kronecker limit formula naturally yields a modular invariant one-loop partition function. The resulting expression generalizes the edge contribution of a massive vector boson in a spontaneously broken gauge theory to the infinite tower in string theory. It is naturally ultraviolet finite and amenable to a state-counting interpretation.

\end{abstract}

\maketitle

\section{Introduction}\label{sec-introduction}

A Euclidean path integral on a manifold $S^1 \times M$ admits a natural Lorentzian interpretation as a thermal partition function. One can regard $M$ as the spatial slice and the generator of time translations along $S^1$ as the canonical Hamiltonian $H$. If $\beta$ is the length of the $S^1$, then the path integral $\hat{Z}(\beta)$ can be identified with the partition function $\Tr_{\mathcal{H}}[e^{-\beta H}]$ where $\mathcal{H}$ is the Hilbert space of wave-functions defined over $M$. A path integral over a generic Euclidean manifold does not admit such a canonical thermal interpretation. 

Surprisingly, the one-loop path integral $\hat{Z}^{i}$ of a quantum field with mass $m_i$ and spin $s_i$ in Euclidean $EdS_{d+1}$ does admit a `nearly thermal' interpretation \cite{Anninos:2020hfj}: 
\begin{equation} \label{bulkedgePI}
	\log \hat{Z}^{i} = \log \hat Z^i_{\text{bulk}} - \log \hat Z^i_{\text{edge}} \, .
\end{equation}
Here $\hat Z^i_{\text{bulk}} = \Tr_S [ e^{-\beta H^i_S}]$, where $H^i_S$ is the Hamiltonian that generates time translations in the dS static patch, and $\Tr_S$ is the trace over the corresponding multi-particle Hilbert space. The correction term $\hat Z^i_{\text{edge}}$ arises from the edge modes localized near the de Sitter horizon. 

The physical origin of this split is the following. One could try to identify the generator of time translations in the static dS patch as the canonical Hamiltonian. Observers in this patch experience a heat bath at inverse temperature $2\pi$, suggesting a thermal interpretation. However, this generator on the Euclidean manifold has a fixed point corresponding to the bifurcate horizon in the Lorentzian continuation. As a result, the spatial slicing is not smooth at the fixed point, leading to a correction term from the modes localized near the horizon. 
 
Both the bulk and the edge contributions can be computed efficiently in terms of the Harish-Chandra characters (which generalize the notion of characters for finite-dimensional representations to infinite-dimensional representations of noncompact groups) of the relevant symmetry group \cite{HarishChandra:1956, HarishChandra:1963}. We would like to use these field-theoretic results near dS horizons to learn about string theory near Rindler horizons. 

Formulating string theory directly in Minkowski-Rindler spacetime and separating the modes localized at the horizon by worldsheet methods is nontrivial because the sigma model for the Rindler wedge is non-Gaussian. One approach for focusing onto horizon modes was proposed in \cite{Dabholkar:1994ai}, and further explored in \cite{Dabholkar:2001if,He:2014gva,Mertens:2015adr,Witten:2018xfj,Dabholkar:2022mxo,Dabholkar:2023ows,Dabholkar:2023yqc,Dabholkar:2023tzd,Dabholkar:2024neq,Dabholkar:2025hri}, using orbifolds of the Euclidean Rindler plane and a stringy generalization of the replica method. The resulting answers are finite both in the UV and the IR. While these results are encouraging, finding a modular invariant expression in closed string theory has proven to be difficult. Other attempts include \cite{Donnelly:2016jet,Balasubramanian:2018axm, Ahmadain:2025pox}. 

In this note, we follow an entirely different approach motivated by the results mentioned above. One can consider the horizon of the dS static patch with a large radius of curvature acting as an infrared regulator. One expects that the near horizon physics is not affected by this regulator. 
Explicit formulae for the splitting between bulk and edge modes are available for fields with arbitrary mass and spin, and therefore, for the entire tower of fields in string theory. We sum the one-loop contribution of all spacetime field modes, following a procedure similar to the one adopted in \cite{Polchinski:1985zf} for the thermal partition function in string theory. 
This naturally leads to an elegant expression for a modular invariant worldsheet partition function for the edge modes. 

To use these field theory results effectively, we compactify 26-dimensional bosonic string theory to three spacetime dimensions with target spacetime of the form $\mathbb{R}^{1,2}\times (S^1)^{23}$, where the $23$ circles all have a common radius $R$. This gives rise to an infinite tower of stringy, as well as Kaluza-Klein, modes with specific masses, spins and multiplicities in $\mathbb{R}^{1,2}$ Minkowski spacetime. One can now replace the Minkowski factor by a $dS_3$ factor with radius of curvature $L$. Even at large $L$, the splitting \eqref{bulkedgePI} allows one to keep the edge modes near the horizon in sight at subleading order in $1/L^2$. There are a number of practical advantages of considering $dS_3$. In $dS_3$, the edge contribution comes only from \textit{massive} states because the only massless propagating states have $s=1$, for which $\chi_{\text{edge}}=0$. Hence, it is sufficient to consider only the principal series of the de Sitter group. Moreover, the little group $SO(2)$ for a massive state is abelian and has only one-dimensional representations, labeled by helicity. This considerably simplifies the group theory. 

One could equally well consider edge modes in $AdS_3$ instead of $dS_3$, which have a very similar structure \cite{Sun:2020ame}. The analysis is somewhat more involved in $AdS_3$ because of the need to regularize the infinite volume, but has the advantage that the string equations of motion can be satisfied. This will be reported in \cite{Dabholkar:2026WiP}. 

One can now define the partition function for the edge modes of all fields in string theory by 
 \begin{equation}\label{EdgeWS}
	Z_{\text{edge}} := \log \hat Z_{\text{edge}} := \sum_i \log \hat Z^i_{\text{edge}} \, ,
 \end{equation}
where we have identified the worldsheet partition function $Z_{\text{edge}}$ with the logarithm of the total spacetime partition function for the edge modes, obtained by summing over all strings fields, extending the usual identification for the bulk modes. We use the `hat' for spacetime partition functions to distinguish from the worldsheet partition functions with no hat. Using the heat kernel representation of the spacetime determinants and identifying the Schwinger parameters with modular parameters, one can obtain a representation of the worldsheet partition function as a modular integral. This is the object that we analyze and show that it naturally leads to a modular invariant expression. 
 
\section{Edge modes in Quantum Field Theory}\label{sec:edge-review}

Consider a bosonic field of mass $m$ and spin $s$ in $dS_3$.
The Hilbert space furnishes a unitary, infinite-dimensional, irreducible representation $[\nu, s]$ of $SO(1, 3)$ with $\nu^2 = L^2 m^2 - (s-1)^2$. The Euclidean continuation of $dS_3$ is $S^3$ with symmetry group $SO(4)$. The one-loop Euclidean partition function is given by
\begin{equation} \label{heatkerneldS}
\begin{split}
	\log \hat{Z} &= - \frac{1}{2} \sum_{n\geq -1} D_{n,s}^{(4)} \log((n+1)^2 + \nu^2) \\
	&= \int_0^\infty \frac{dt}{2t} e^{-t \nu^2} \sum_{n\geq -1} D_{n,s}^{(4)} e^{-t \left( n + 1\right)^2 } \, , \\
\end{split}
\end{equation}
where $D_{n,s}^{(4)}$ are the $SO(4)$ multiplicities and $(n+1)^2$ is related to the eigenvalue of the Laplacian by a shift \cite{David:2009xg, Anninos:2020hfj}. It admits a Lorentzian interpretation \cite{Anninos:2020hfj}:
\begin{equation} \label{oneloopchar}
	\log \hat{Z} = \int_0^\infty \frac{du}{2u} \frac{1+ e^{-u}}{1 - e^{-u}} \, \left(\chi^{\text{dS}}_{\text{bulk}}(u) - \chi^{\text{dS}}_{\text{edge}}(u) \right) \, ,
\end{equation}
where $\chi^{\text{dS}}_{\text{bulk}}(u)$ is the Harish-Chandra character \cite{HarishChandra:1956, HarishChandra:1963} of the representation $[\nu, s]$:
\begin{equation}
	\chi^{\text{dS}}_{\text{bulk}}(u) = \chi(u) \equiv \tr_G e^{-i u H} \, ,
\end{equation}
where $H$ is the global $SO(1,1)$ generator, which acts as the timelike Killing flow in the static patch, and $\tr_G$ is the trace over the global single-particle Hilbert space that furnishes the representation. The equivalence between the trace over single-particle states in the static patch and $\tr_G$ is a consequence of a one-to-one mapping between states with the same $\omega$ in the static and global patches via the Bogoliubov transformation. For a massive integer spin-$s \geq 1$ field, the bulk Harish-Chandra character is given by
\begin{equation}\label{bulk-char}
	\chi^{\text{dS}}_{\text{bulk}}(u) = D_s^{(2)} \,
	\frac{ e^{- u (1 + i\nu) } + e^{- u (1 - i\nu )} }{ (1 - e^{-u})^{2} } \, , 
\end{equation}
where $D^{(d)}_s$ is the $SO(d)$ spin degeneracy. In particular, $D^{(2)}_s =2$, since all $SO(2)$ representations are one-dimensional and a spin $s$ field contains the contributions from both helicities $\pm s$. Note also that $D^{(4)}_{s} = (s+1)^2$.

Interestingly, the edge contribution also turns out to be given in terms of a Harish-Chandra character, but in two fewer dimensions \cite{Anninos:2020hfj}:
\begin{equation} \label{chiedge}
	\chi^{\text{dS}}_{\text{edge}}(u) = D^{(4)}_{s-1} \,
	\left( e^{- i u\nu } + e^{ i u \nu} \right) \, ,
\end{equation}
As shown in \cite{Anninos:2020hfj}, integrating over $\chi^{\text{dS}}_{\text{bulk}}(u)$ as in \eqref{oneloopchar} gives $ \log \hat Z^i_\text{bulk}$ which equals the logarithm of the thermal partition function. Thus, it is clear from \eqref{oneloopchar} that for a generic bosonic field with spin, the Euclidean path integral is not exactly equal to the thermal partition function. A correction term is needed, which can be identified with the contribution from the edge modes, $\log \hat{Z}^i_{\text{edge}}$ in \eqref{bulkedgePI}. 

The representation above in terms of the characters is useful to see the split between the bulk and the edge modes from the Lorentzian perspective. This split can also be seen from the Euclidean perspective \cite{Anninos:2020hfj}, by noting that for $n\geq 0$, 
\begin{equation}\label{EuclideanSplit}
\begin{split}
	D^{(4)}_{n,s} &= D^{(2)}_s D^{(4)}_n - D^{(4)}_{s-1} D^{(2)}_{n+1} \\
	&=2 ( n+ 1)^2 - 2 s^2 \, , 
\end{split}
\end{equation}
while $D^{(4)}_{-1,s} = -s^2$. Substituting into \eqref{heatkerneldS}, the $2(n+1)^2$ term gives rise to the bulk character, and the $s^2$ terms to the edge character, which equals the character of $s^2$ scalar fields in two lower dimensions. This group theoretic formulation has a physical interpretation in terms of quasi-normal modes \cite{Anninos:2020hfj,Sun:2020sgn}, which could be useful in less symmetric backgrounds \cite{Law:2022zdq,Grewal:2022hlo}. 

The formal expression \eqref{oneloopchar} diverges for small $u$, corresponding to the UV divergence in QFT. Our goal is to examine the entire string tower of edge modes to obtain a modular invariant expression. The UV divergence is then naturally regularized by the string length $\ell$ by restricting the Schwinger integral to the fundamental domain. 

\section{Edge Modes in String Theory} \label{sec-horizon-bosonic}

We first discuss the bulk partition function arising from the first term in \eqref{EuclideanSplit}, common to all modes. For large $L$, $dS_3$ can be replaced by $\mathbb{R}^{1,2}$ with $L$ as a cutoff. Adding the $(S^1)^{23}$ factor and summing over all states as in \cite{Polchinski:1985zf} one expects to obtain the one-loop cosmological constant going as $L ^3 R^{23}$. Substituting $t = \pi \ell^2 \tau_2/L^2$ in \eqref{heatkerneldS}, 
\begin{equation}
\begin{split}
	\log\hat Z_{\text{bulk}}^i = \int_0^\infty \frac{d\tau_2}{\tau_2} \sum_{n\geq -1} 
	( n+ 1)^2 e^{- \frac{ \pi \ell^2 \tau_2}{L^2} \left( \left( n + 1\right)^2 + \nu_i^2 \right)} \, . 
\end{split}
\end{equation}
Replacing $n+1= L \rho $, taking the large $L$ limit, and performing the $\rho$ integral, we obtain
\begin{equation} \label{flatspdS}
\begin{split}
	\log \hat{Z}_{\text{bulk}}^i &\sim \frac{L^3}{\ell^3 }
	\int_0^\infty \frac{d\tau_2}{\tau_2^{5/2}} e^{- \pi \ell^2 \tau_2 m_i^2 } \, .
\end{split}
\end{equation}
As in \eqref{EdgeWS}, one can define the bulk worldsheet partition function as a sum over particle species labeled by $i$:
\begin{equation}
	Z_{\text{bulk}} := \log \hat{Z}_{\text{bulk}} := \sum_i \log \hat{Z}_{\text{bulk}}^i \, . 
\end{equation}
Therefore, 
\begin{equation}
	\begin{split}
	Z_{\text{bulk}} \sim \frac{L^{3}}{ \ell^{ 3} } \int_0^\infty \frac{d\tau_2}{\tau_2^{5/2}} \sum_i e^{ - \pi \ell^2\tau_2 m_i^2} \, .
	\end{split}
\end{equation}
Masses of states labeled by $i$ in the spectrum of bosonic string theory are given by the on-shell condition $\frac{1}{2} \ell^2 m_i^2= N + \tilde N - 2 + \frac{1}{2R^2} \sum_{r=1}^{23} n_r^2 \ell^2$, where $N$ and $\tilde N$ are left and right moving oscillator energies respectively, and $\{n_r\} $ are the Kaluza-Klein momenta around the $23$ circles. One must also impose level matching by inserting $\delta_{N,\tilde N} = \int d\tau_1 \, e^{2 \pi i \tau_1 (N - \tilde N) }$. Thus, the sum over $i$ turns into a worldsheet trace $\tr$ over the oscillator modes of the string in flat space, as in \cite{Polchinski:1985zf}. For large $R$, the sum over Kaluza-Klein momenta turns into an integral which yields a factor of $R^{23}/\ell^{23}\tau_2^{23/2}$ to give
\begin{equation}
 	Z_{\text{bulk}} = \frac{L^{3}R^{23}}{ \ell^{ 26} } \int_\mathcal{S} \frac{d^2\tau}{\tau_2^{14}} \tr{}\,\left[ q^{ N-1} \bar{q}^{\tilde{N} -1} \right]\, ,
\end{equation}
with $\tau:= \tau_1 + i \tau_2$ and $q:= \exp{(2\pi i \tau)}$. The sum over oscillator states gives the worldsheet partition function:
\begin{equation}\label{spacetime-bosonic}
	Z_{\text{bulk}} \sim \frac{L^{3}R^{23}}{\ell^{26}} \int_\mathcal{S} \frac{d^2\tau}{\tau_2^2}\, \Lambda(\tau) \, , 
\end{equation}
where, as expected, 
\begin{equation}\label{cosmo-density}
	\Lambda(\tau):= \frac{1}{\tau_2^{12}} \frac{1}{|\eta(\t)|^{48}} 
\end{equation}
is the one-loop cosmological constant density of the bosonic string in $\mathbb{R}^{1, 25}$. The Weil-Petersson measure $d^2\tau/\t_2^2$ of the integral and $\Lambda(\tau)$ are manifestly modular invariant under modular transformations
\begin{equation} \renewcommand{\arraystretch}{0.9}
	\gamma \left( \tau \right) =\dfrac{a\tau+b}{c\tau+d}\, , \quad \gamma = 
	\left(\begin{array}{cc}
	a & b \\
	c & d\\
	\end{array}\right)
	\in \Gamma \equiv SL\left( 2,\mathbb{Z} \right) \, . 
\end{equation} 
The integration domain in quantum field theory is the strip $\mathcal{S}$ ( $-\frac{1}{2} \leq \tau_1 < \frac{1}{2} $ and $\tau_2 \geq 0$) in the upper-half $\tau$ plane $\mathbb{H}$ ($\tau_2 \geq 0$). It is well-known that in string theory, the integration domain is restricted to the `key-hole' fundamental domain $\mathcal{F} := \Gamma\backslash\mathbb{H}$ of the modular group, which automatically excludes the $\tau_2 \rightarrow 0$ region responsible for the UV divergences in quantum field theory. This is possible because of modular invariance. 
 
We now turn to the $s^2$ term, corresponding to the edge terms. Summing over particle species gives
\begin{equation}\label{sum-spin}
	\begin{split}
	Z_{\text{edge}} \sim - \frac{L\,R^{23}}{ \ell^{ 24} } \int_0^\infty \frac{d\tau_2}{\tau_2^{13}} \sum_i \, s_i^2 \, e^{ - \pi \ell^2\tau_2 m_i^2} \, .
	\end{split}
\end{equation}
In the flat limit, we have three real worldsheet bosons with $\mathbb{R}^3$ as the target space. We combine two of them into a complex boson $X:= X_1 + i X_2$ with unit charge under the $U(1)$ generator, corresponding to the rotation in the plane. The remaining direction $X_3$ is uncharged. The quantum number $s$ corresponds to the spin in the 1-2 plane, excluding the orbital angular momentum coming from the zero modes. Summing over species in \eqref{sum-spin} implies inserting an operator $\half (S + \tilde{S})^2$ in the worldsheet trace, where $S$ and $\tilde S$ are the spin operators of the left and right movers. The factor of half accounts for the fact we sum over both $\pm s$ helicities.

Thus, the worldsheet partition function \eqref{sum-spin} for the edge modes on the horizon takes the form
\begin{equation}\label{edge-one}
	Z_{\text{edge}} \sim -\frac{L\,R^{23}}{ \ell^{ 24} } \int_0^\infty \frac{d^2\tau}{\tau_2^{13}} \tr{}\,\left[ (S+ \tilde{S})^2 \, q^{ N-1} \bar{q}^{\tilde{N} -1} \right]\, .
\end{equation}
The part of the trace coming from the oscillators of the complex boson (including the vacuum energy) gives
\begin{equation}
\begin{split}
	& \, \tr_X \left[ (S + \tilde{S})^2 \, q^{ N-1/12} \bar{q}^{\tilde{N} -1/12} \right] \\
	&= \lim_{\mu\rightarrow 0} \; \frac{1}{4\pi^2 } \frac{\partial^2}{\partial \mu^2} \; Z^X(\tau,\mu) \overline{Z^X(\tau, \mu)} \, , \\
\end{split}
\end{equation}
where 
\begin{equation}
\begin{split}
	Z^X(\tau,\mu) &= \tr_X \left( q^{N - \frac{1}{12} } e^{2\pi \mu S }\right) \, ,
	\end{split}
\end{equation}
with $z= -\bar{z}= i \mu$, where $\mu$ is a real chemical potential. The trace over left-movers gives
\begin{eqnarray}
	Z^X(\tau,\mu) &=& q^{-\frac{1}{12}}\prod_{n=1}^{\infty}
	\frac{1}{(1 - q^n e^{2\pi \mu})(1 - q^n e^{-2\pi \mu})} \, \nonumber\\
	&=&\frac{2 \sin(\pi i \mu)\, \eta(\tau) }
	{\vartheta_1(i \mu|\tau)} \, , 
\end{eqnarray}
where $\eta(\tau)$ is the Dedekind function and $\vartheta_1(z|\tau)$ is the Jacobi function. Expanding near $\mu = 0$,
\begin{equation}
\begin{split}
	\frac{\sin(\pi i\mu) }
	 {\vartheta_1(i\mu|\tau)} 
	\approx \frac{1}{2 \eta(\tau)^3 } \left( 1 + \frac{ \pi^2 \mu^2}{6}\left( 1 - E_2(\tau) \right) + \ldots \right) \, . 
\end{split}
\end{equation}
Evaluating $\frac{1}{4\pi^2 } \frac{\partial^2}{\partial \mu^2} \left| Z^X(\tau,\mu)\right|^2 \bigg|_{\mu=0}$ one obtains
\begin{equation} \label{E2expression}
\begin{split}
	\frac{1}{12} \frac{1}{ |\eta(\tau)|^4 } \left( 2 - E_2(\tau) - \overline{E_2(\tau)} \right) \, . 
\end{split}
\end{equation}
Here $E_2(\tau)$ is the quasimodular form defined by
\begin{equation}\label{E2}
	E_2(\tau) = \frac{1}{2\pi i }\frac{d}{d\tau} \log [\eta(\tau)^{24}] = 1 -24 \sum_{n=1}^\infty \frac{n q^n}{ 1-q^n} \, .
\end{equation}
Ignoring the constant term, which we comment upon later, the edge mode partition function is given by
\begin{equation}
	\begin{split}
	Z_{\text{edge}} \sim & \frac{L\,R^{23}}{ \ell^{ 24} } \int_{\mathcal{S}} \frac{d^2\tau}{\tau_2^2}\, \Lambda (\tau) .
	\end{split} \left( \t_2 E_2(\tau) + \t_2 \overline{E_2(\tau)}\right) \, . 
\end{equation}
Since the edge modes live on a co-dimension two surface with two fewer momentum integrals, there is a factor of $\tau_2$ relative to the bulk contribution. The terms in the bracket can be written as
\begin{equation}
	\left( \tau_2 \hat E_2(\tau) + \overline{\tau_2 \hat E_2(\tau) } + \frac{6}{\pi }\right) \, , 
\end{equation}
where $\hat E_2(\t) = E_2(\t) -3/{\pi\t_2}$ transforms as a weight $+2$ modular form. The constant term $6/\pi$ gives a subleading correction to the cosmological constant, which we ignore. 

The strip $\mathcal{S}$ can be mapped to the fundamental domain $\mathcal{F}$ by modular transformations $\g$ in the coset $ \Gamma_\infty\backslash\Gamma$ where $\Gamma_\infty = \left\{
\pm \left( \begin{smallmatrix}
	1 & n \\ 
	0 & 1 \, 
\end{smallmatrix}\right) \, , \, n \in \mathbb{Z} \right\}$ is the subgroup of translations fixing $\infty$. These elements are in one-to-one correspondence with a pair of coprime integers $(c,d)$. Thus, 
\begin{equation}
	\begin{split}
	Z_{\text{edge}} \sim & \frac{L\,R^{23}}{ \ell^{ 24} } \int_{\mathcal{F}} \frac{d^2\tau}{\tau_2^2}\, \Lambda (\tau) .
	\end{split} \left( F(\tau) \, \hat E_2(\tau) + c. c. \right) \, , 
\end{equation} 
where 
\begin{equation}\label{Fdef}
	F(\tau) = \, \frac{1}{2} \tau_2 
	\sum_{\substack{c,d \in \mathbb{Z}\\(c,d)=1}}
	\dfrac{(c\tau + d)^2} {|c\tau + d|^2} \, ,
\end{equation}
which formally transforms as a modular form of weight $-2$, and the integrand is modular invariant. However, the sum in \eqref{Fdef} is not convergent. We therefore define $F(\tau)$ as a limit ${s \rightarrow 0}$ of a modular covariant function
\begin{equation}
	F(\tau, s) = \, \frac{1}{2} \tau_2 
	\sum_{(c,d)=1}
	\dfrac{(c\tau + d)^2} {|c\tau + d|^2}
	\dfrac{\tau_2^s}{|c\tau + d|^{2s}} \, .
\end{equation}
This can be viewed as a modular invariant version of dimensional regularization since varying dimensions in the momentum integrals changes powers of $\tau_2$ appearing in the Schwinger integral. The function $F(\tau, s)$ transforms like a holomorphic modular form with weight $-2$:
\begin{eqnarray}
	F(\gamma(\t), s) = (c\tau + d)^{-2} F(\t, s) \, , \qquad \forall \,\gamma \in \Gamma \, .
\end{eqnarray}
We note that 
\begin{equation} \label{Maass}
	F(\t, s) = -2 i \t_2^2 \frac{\partial}{\partial \bar{\tau}} \left[ \frac{1}{s} E\left( \tau, s \right) \right]\, .
\end{equation}
The differential operator on the right is the Maass lowering operator \cite{Osullivan:2018}, which reduces the modular weight by $2$. It acts on the non-holomorphic Eisenstein series 
 \begin{equation}
	E\left( \tau, s \right) := \langle\tau_2^s\rangle = \frac{1}{2}
	\sum_{(c,d)=1}\dfrac{\tau _{2}^{s}}{| c\tau+d| ^{2s}} \,
	\, .
\end{equation}
The sum is absolutely convergent for $\Re (s) > 1 $ and manifestly modular invariant:
\begin{equation}
	E\left( \gamma ( \tau ), s \right) = E\left( \tau , s \right) \, , \quad \forall \,\gamma \in \Gamma \equiv SL\left( 2,\mathbb{Z} \right) \, .
\end{equation}
It can be shown that the function defined by
\begin{equation}
	E^*\left( \tau, s \right) := \pi ^{-s}\Gamma\left(s \right) \zeta \left( 2s \right) \, E\left( \tau, s \right) \, ,
\end{equation}
with Euler Gamma function $\Gamma$ and Riemann zeta-function $\zeta$, can be analytically continued to the complex $s$ plane with poles only at $s=0$ and $s=1$ \cite{Zagier:1981}, satisfying 
\begin{equation}
	E^*\left( \tau ,s\right) =E^*\left( \tau ,1-s\right) \, .
\end{equation} 
It implies that $E(\tau, s) \sim s E(\tau, 1-s) $ near $s=0$. Thus, instead of $s= 0^+$ we can examine the behavior near $s=1^-$. Precisely this Laurent expansion is given by the classic Kronecker limit formula \cite{Lang1987}:
\begin{equation}\label{Kronecker}
	E(\tau, s) = \frac{\pi}{s-1} - \pi\log (\tau_2 |\eta(\tau)|^4) + c + \mathcal{O}(s-1) \, , 
\end{equation}
where $c$ is a constant. It is nontrivial that the residue at the pole at $s=1$ is independent of $\tau$, and thus acting with the Maass lowering operator in \eqref{Maass}, and taking $s \rightarrow 1$, one obtains a finite expression:
\begin{equation}
	F(\tau) = - \frac{\pi^3}{9}\tau_2^2 \overline{\hat{E}_2(\tau)} \, , 
\end{equation}
which is a modular form of weight $-2$. The resulting edge partition function is manifestly modular invariant:
\begin{equation}\label{hor}
	Z_{\text{edge}} \sim \frac{L\,R^{23}}{ \ell^{ 24} } \int_{\mathcal{F}} \frac{d^2\tau}{\tau_2^2}\, \Lambda (\tau) \left(\tau_2^2 \hat E_2(\tau) \overline{ \hat E_2(\tau) }\right) \, .
\end{equation} 
Since $E_2 (\tau)$ admits a $q$-expansion, \eqref{E2}, $Z_{\text{edge}}$ in \eqref{hor} seems amenable to a state-counting interpretation consistent with the spacetime interpretation of \eqref{chiedge} as the contribution of $D^{(4)}_{s-1}=s^2$ field degrees of freedom in two fewer dimensions. It would be interesting to find a direct worldsheet evaluation. 

The constant term in \eqref{E2expression}, multiplied by $\tau_2$, can similarly be analytically continued and gives a modular invariant expression, but now proportional to $E(\tau, s)$ near $s=1$, which has a pole at $s=1$ \eqref{Kronecker}. We do not know how to interpret this unphysical divergence. One possibility is that a subleading contribution to the central charge cancels this constant term because de Sitter space is slightly off-shell and does not satisfy the string equations of motion in spacetime at subleading order in $L$. Since the spacetime action is related to the central charge, such a contribution is plausible but it remains to be seen if it has the right coefficient. 

\section{Physical Interpretation} \label{sec:discussion}

Edge modes arise in gauge theory due to gauge constraints, such as Gauss' law for a Maxwell field \cite{Donnelly:2011hn,Donnelly:2014gva,Donnelly:2014fua,Donnelly:2015hxa,Ball:2024hqe}. For a Proca field in a spontaneously broken gauge theory, the necessity for edge modes can be seen most clearly by quantizing the Fierz-Pauli system \cite{Blommaert:2018rsf, Anninos:2020hfj}. String theory has a large (possibly spontaneously broken) gauge invariance, with an infinite tower of massive fields. The partition function \eqref{hor} captures the edge modes for this tower. For each mass $m$ and helicity $r$, one obtains a unitary irreducible representation $[m, r]$ of the Poincar\'e group $ISO(1,2)$. CPT invariance requires that, for a given spin $s$, both helicities $r= \pm s$ must belong to the spectrum. The representation $[m_i ,s_i]\oplus [m_i , -s_i]$ can be obtained as a Hilbert space by quantizing a Fierz-Pauli theory with St\"uckelberg fields for a totally symmetric tensor $\Phi_{\m_1\mu_2 \ldots \m_s}$ which is transverse and traceless ($\nabla^\nu \Phi_{\nu\m_1\ldots \m_{s-1}} = \Phi^\nu_{\nu\m_1\ldots \m_{s-2}} =0$), described by a local gauge-invariant action. 

Instead of computing the edge contributions near the Rindler horizon directly, we have considered its deformation to the de Sitter horizon with extremely low curvature. The states now furnish representations $[\nu_i ,s_i]\oplus [\nu_i , -s_i]$ in the principal series of the de Sitter group $SO( 1, 3)$, regarding the Poincar\'e group as a Wigner-In\"on\"u contraction of the de Sitter group. Properly taking into account the contribution from the `unmatched' zero modes of the Faddeev-Popov ghost fields, the bulk and edge contributions organize into Harish-Chandra characters \eqref{oneloopchar} consistent with locality and unitarity \cite{Blommaert:2018rsf, Anninos:2020hfj}.

A short proper time expansion of the character integral \eqref{oneloopchar} is in agreement with the known heat kernel coefficients \cite{Vassilevich:2003xt}. The edge corrections could be interpreted as `contact terms', as suggested by the heuristic picture of Susskind and Uglum \cite{Susskind:1994sm} of a string worldsheet at the horizon and computed for the photon and the graviton field in \cite{Kabat:1995eq,David:2022jfd,Blommaert:2024cal}. The contact terms arise from couplings linear in the Riemann tensor in the quadratic fluctuation operator and are generically present for any spin-$s\geq1$ field. We emphasize that the modular invariant expression \eqref{hor} is free of UV divergences. Moreover, the integral over the proper time contains a lot more information than the short proper-time expansion, including the nonlocal and finite contributions from the infinite tower of Kaluza-Klein modes in the large $R$ limit. 

The edge modes contribute to the entanglement entropy associated with the horizon as the entangling surface. They could also be related to soft modes \cite{Anninos:2021ihe}. The formal expression \eqref{oneloopchar} for $\hat{Z}_{\text{edge}}$ in quantum field theory is UV divergent. As we have seen, after summing over the string tower, one obtains a UV convergent expression \eqref{hor} because of modular invariance. For the bosonic string there is the usual tachyonic IR divergence, which is expected to be absent for the superstring. Moreover, one expects that the total entanglement entropy will receive contributions also from the bulk term, corresponding to the area divergence due to the divergent local temperature of the thermal bath near the horizon. These questions will be explored in \cite{Dabholkar:2026WiP}. Independently of its relation to entanglement entropy, $Z_{\text{edge}}$ is interesting in its own right, since it is related to the underlying gauge symmetry. 

It is useful to formulate these results in the language of algebraic quantum field theory. In this formalism there is no UV cut-off. The UV divergence of $\hat Z_{\text{edge}}$ (or of entanglement entropy) corresponds to the fact that, unlike in a bipartite quantum mechanical system, the algebra of observables in a QFT is Type-III rather than Type-I. See \cite{Witten:2018zxz} for a review. With the inclusion of gravity, the algebra of observables can become Type-II \cite{Witten:2021unn, Chandrasekaran:2022cip}, a fact possibly related to the renormalization of Newton's constant in canonical gravity \cite{Susskind:1994sm}. However, one would really like to show that the algebra is more like Type-I. This is expected also from the holographic description of the two-sided black hole, since the algebra of observables acting on a single copy of the dual CFT furnishes an irreducible representation on the Hilbert space of the CFT, and is thus Type-I. It is not clear how to define local observables in string theory but the UV finiteness of $Z_{\text{edge}}$ is a strong indication of a Type-I behavior. 

In QFT with a lattice UV cutoff, the algebra of observables can have a `center' due to the gauge constraints \cite{Buividovich:2008gq,Casini:2013rba,Ghosh:2015iwa,Moitra:2018lxn}. Edge modes near the entangling Rindler horizon are a reflection of the fact that the Hilbert space does not factorize between `inside' and `outside' of the horizon. In string theory, the string length acts as an effective UV cutoff, and hence one expects that the partition function $Z_{\text{edge}}$ will contain information about this center of the algebra for the entire stringy tower. It would be interesting to see if the string replica method \cite{Dabholkar:1994ai} with a suitable analytic continuation \cite{Dabholkar:2023ows} can reproduce these results; see \cite{Dabholkar:2024neq} for related comments. 

\newpage
\vspace{1em}
\centerline{\textit{Acknowledgements}}
\vspace{1em}
We would like to thank Dionysios Anninos, Valentin Benedetti, Zimo Sun, and especially Caner Nazaroglu for useful discussions. UM is supported by the European Research Council (ERC) under the European Union’s Horizon 2020 Research and Innovation Programme (Grant Agreement No. 101115511).


%

\end{document}